\documentclass{JHEP3} 

\usepackage{amsmath,amssymb}
\usepackage{graphicx}
\usepackage[latin1]{inputenc}
\newcommand{\onecol}[2]{
        \begin{minipage}[t]{#1}{#2\vfill} \end{minipage}
        }

\newcommand{\psibar}{\overline{\psi}}
\newcommand{\psihat}{\hat{\psi}}

\newcommand{\zetabar}{\overline{\zeta}}

\newcommand{\gmu}{\gamma_\mu}

\newcommand{\go}{\gamma_0}

\newcommand{\gfive}{\gamma_5}
\newcommand{\dmu}{\partial_\mu}
\newcommand{\dmub}{{\partial^\ast}_{\mspace{-11.0mu} \mu}}
\newcommand{\db}[1]{{\partial^\ast}_{\mspace{-11.0mu} #1}}

\newcommand{\md}{\widetilde{\partial}}

\newcommand{\e}{\mathrm{e}}

\newcommand{\rO}{\mathrm{O}}
\newcommand{\<}{\langle}
\newcommand{\z}{\rangle}

\newcommand{\li}{\left}
\newcommand{\re}{\right}
\newcommand{\ren}[1]{\li(#1\re)_\mathrm{R}}
\newcommand{\ev}[1]{\bigl\<#1\bigr\z}
\newcommand{\threevector}[1]{\mathbf{{#1}}}
\newcommand{\vx}{\threevector{x}}

\newcommand{\vp}{\threevector{p}}
\newcommand{\phat}{\hat{p}}
\newcommand{\vphat}{\hat{\vp}}

\newcommand{\Dw}{D_\text{W}}

\newcommand{\eqw}[1]{\stackrel{#1}{=}}
\newcommand{\Op}{\mathcal{O}}

\newcommand{\Oss}{O_{SS}}
\newcommand{\Opp}{O_{PP}}
\newcommand{\Ovv}{O_{VV}}

\newcommand{\Ohss}{\hat{O}_{SS}}
\newcommand{\Ohpp}{\hat{O}_{PP}}
\newcommand{\Ohvv}{\hat{O}_{VV}}

\title{Test of the Schr\"odinger functional with chiral fermions in the
Gross-Neveu model}

\author{Bj\"orn Leder\\
        Trinity College Dublin,\\
        College Green, Dublin, Ireland\\
        E-mail: \email{leder@maths.tcd.ie}
        \hfill
        \onecol{3cm}{\vspace{-4em}\it
            TCDMATH 07--20\\
            TRINLAT-07/05}
         }

\abstract{The recently proposed construction of chiral fermions on lattices with
boundaries is tested in an interacting theory up to first order of perturbation
theory. We confirm that, in the bulk of the lattice, the chiral Ward identities
take their continuum value up to cutoff effects without any tuning. Universal
quantities are defined that have an expansion in the renormalised couplings with
coefficients that are functions of the physical size and the periodicity in the
spatial direction. These coefficient functions have to be identical for
different discretisations. We find agreement with the standard Wilson fermions.
The computation is done in the asymptotically free Gross-Neveu model with
continuous chiral symmetry.}


\begin{document}

\section{Introduction}
We present a 1-loop perturbative calculation that confirms the
favourable properties of chiral fermions on a lattice with boundaries
\cite{Luscher:2006df}.
The calculation is performed in the chiral Gross-Neveu model
\cite{Gross:1974jv}. Before I
go into the details and results of the calculation let me briefly introduce
field theories on lattices with boundaries.

In the Schr\"odinger functional (SF) of a quantum field theory the fields are
defined on a $d+1$ dimensional cylinder. The fields are subject to Dirichlet
boundary conditions in the time direction ($P_{\pm}=\tfrac{1}{2}(1\pm \go)$)
\begin{equation}\label{bc_1}
   P_+ \psi(x) = 0\,,\quad \psibar(x) P_- = 0 \quad \text{at} \quad
x_0=0\,, 
\end{equation}
\begin{equation}\label{bc_2}
   P_- \psi(x) = 0\,,\quad \psibar(x) P_+ = 0 \quad \text{at} \quad x_0=T
\end{equation}
and periodic boundary conditions in the space directions
\begin{equation}\label{bc_3}
\psi(x+L\hat{k})=\e^{i\theta}\psi(x)\,,\quad \theta\in[0,2\pi)\,, \quad
      k=1,\dots d\,.
\end{equation}

Discretising such a theory on a space-time lattice with
boundaries has advantages. If the temporal extension is a multiple of the
spatial extension, say $T=2L$, the inverse of the spatial extension $1/L$
provides an infrared
cutoff and is the natural energy scale in the massless theory. In fact this has
enabled a fully non-perturbative determination of the scale dependence of the
fundamental parameters of $N_{\mathrm{f}}=2$ QCD, see \cite{Sommer:2006wx} and
references therein.

However, naively one would expect that theories become more complicated on
lattices with boundaries. First of all the distinction between
different classes of boundary conditions, for example Dirichlet and Neuberger
boundary conditions, really only makes sense for smooth fields, which
we do not have on the lattice. The boundary conditions therefore must be
encoded in the lattice action and will arise dynamically in the continuum limit.
In general this might involve the necessity to fine-tune some parameter in the
action. Secondly the boundaries might cause additional divergences and thus
lead to additional counter-terms.
Both of these issues have been addressed for QCD in \cite{Sint:1993un} with
the
result that there is no need for fine-tuning and that one just has to add a
renormalisation factor for the boundary fields.

The fact that the SF boundary
conditions arise naturally (without fine-tuning) in the continuum limit has
been understood on the basis of dimensional counting and boundary critical
phenomena, see \cite{Luscher:2006df} and references therein. The SF boundary
conditions are
part of the definition of the continuum limit. Together
with the symmetry properties and the dimensionality of the system they form a
SF universality class: any local discretisation has this continuum
limit.\footnote{Note that in \cite{Sint:2005qz} a SF is proposed with chirally
rotated boundary conditions, which however break parity and
therefore are distinct from the boundary conditions \eqref{bc_1} considered
here.}

Exactly this observation led the author of \cite{Luscher:2006df} to a
formulation of the 
lattice SF of QCD with chiral fermions. In the continuum the SF boundary
conditions break chiral symmetry. Therefore lattice fermions that fulfil the
Ginsparg-Wilson relation \cite{Ginsparg:1981bj} on the whole lattice (with
boundaries) can not
have the right continuum limit. The minimal modification to the Ginsparg-Wilson
relation that breaks the lattice chiral symmetry at the boundary would
be a term $\Delta_B$ that is supported at the boundary with exponentially
decaying tails
\begin{equation}\label{mod_GW}
 \gfive\,D + D\,\gfive = a\,D\gfive D + \Delta_B\,.
\end{equation} 
Given a local solution to this equation chiral Ward identities are expected
to take their continuum form far away from the boundaries.

As mentioned above, in principle there are many possible local lattice
formulations that have the same continuum limit. In \cite{Luscher:2006df} the
author gives one
solution to the modified Ginsparg-Wilson relation, a modified Neuberger-Dirac
operator
\begin{equation}\label{sf_neuberger}
   D=\tfrac{1}{\bar{a}}\li\{1-\tfrac{1}{2}(U+\gfive
               U^\dagger\gfive)\re\}\,,
\end{equation} 
\begin{equation} 
   U=A\,(A^\dagger A + caP)^{-1/2}\,,\quad
   \bar{a}=\frac{a}{1+s}\,,\quad A = 1 + s - a\Dw\,,
\end{equation} 
where $\Dw$ is the massless Wilson-Dirac operator in the presence of the
boundaries as introduced in \cite{Sint:1993un}, i.e. the standard
infinite lattice Wilson-Dirac operator in the range $0<x_0<T$ and at all other
times the target field $\chi=D_W\psi$ is set to zero.\footnote{This Dirac
operator maps the space of fields defined at all $x_0$, but set to zero at
$x_0<a$ and $x_0>T-a$, into itself. }
The projector $P$ is defined through
\begin{equation}
   P\psi(x) = \frac{1}{a}\li\{\delta_{x_0,a}P_-\psi(x)\big|_{x_0=a} +
      \delta_{x_0,T-a}P_+\psi(x)\big|_{x_0=T-a}\re\}\,,
\end{equation}
and the parameters $c$ and $s$ must be chosen such as to ensure locality
(see next section).
The operator \eqref{sf_neuberger} solves \eqref{mod_GW} if the lattice 
spacing $a$ is replaced by the rescaled value $\bar{a}$. For all unexplained
notation we refer the reader to \cite{Leder:2007aq}.

\section{Free fermions}
For Wilson fermions the free propagator in presence of the boundaries can be
calculated explicitly and is given in \cite{Luscher:1996sc}. In the case of
chiral fermions
we do not even have an analytic expression for the Dirac operator. To obtain
the free propagator in closed form like in the Wilson case might be possible but
is much more difficult. Nevertheless, the operator under the square root in
\eqref{sf_neuberger} can be worked out in the time-momentum representation.

Taking into account the definition of $\Dw$ in the presence of the boundaries
given above, the operator under the square root explicitly
reads
\begin{equation}\label{AAdagger}
   A^\dagger A + caP = (1+s)^2 + s a^2 \sum_\mu \dmub\dmu +
      \tfrac{1}{2}a^4\sum_{\mu<\nu} \dmub\dmu\db{\nu}\partial_\nu + (c-1)aP\,,
\end{equation}
with forward $\dmu\psi(x)=\tfrac{1}{a}(\psi(x+a\hat{\mu})-\psi(x))$ and backward
$\dmub\psi(x)=\tfrac{1}{a}(\psi(x)-\psi(x-a\hat{\mu}))$ finite
differences. It is Hermitian and therefore has real eigenvalues. Due
to translation invariance in space the spatial eigenfunctions are plane waves
with momentum values that are integer multiples of $2\pi/L$ in the range
$-\pi/a<p_k\le\pi/a$.\footnote{Greek indices $\mu,\nu,\dots$ run from
$0$ to $d$ and Latin indices $k,l,\dots$ from $1$ to $d$.} For $c=1$ the
operator \eqref{AAdagger} is diagonal in 
Dirac space, so its eigenvalues are $2^{(d+1)/2}$-fold degenerated. Its
eigenfunctions (in space-time) are then given by
\begin{equation}\label{c_one}
   c=1:\quad v_{p_0,\vp}(x)=\e^{i\vp\vx}\,\sin(p_0 x_0)\,, \quad p_0 =
\frac{n\pi}{T}\,,\quad n=1,2,\dots,T/a-1\,.
\end{equation}
For $c\neq1$ the operator \eqref{AAdagger} is not diagonal in Dirac space. The
eigenfunctions are still given in terms of $\sin$-functions
\begin{equation}\label{eigenfunctions}
   c\neq 1:\quad w_{p_0,\vp}(x)=\e^{i\vp\vx}\,
   \li\{  P_-\sin\big(p_0 x_0 + b(p_0)\big)
        + P_+\sin\big(p_0(T-x_0) + b(p_0)\big)\re\}\,,
\end{equation}
where
\begin{equation}
   b(p_0)=-\arctan\li(\frac{\sin(ap_0)}{\frac{q}{c-1}+\cos(ap_0)}\re)\,, \quad
            q=\tfrac{a^2}{2}\vphat^2 - s\,, \quad \phat_\mu =
\tfrac{2}{a}\sin(ap_\mu/2)\,.
\end{equation}
But the allowed values of $p_0$ are now given by the solutions of the
transcendental equation
\begin{equation}\label{p0_equation}
   b(p_0)=-p_0 T\,.
\end{equation}
In either case the eigenvalues are given by
\begin{equation}\label{eigenvalues}
   \lambda_{p_0,\vp} = (1+s)^2 - sa^2\phat^2 +
   \tfrac{a^4}{2}\sum_{\mu<\nu}\phat_\mu^2\phat_\nu^2\,.
\end{equation}
Rewriting this as
\begin{equation}\label{eigenvalues2}
   \lambda_{p_0,\vp} = q\,a^2\phat_0^2 + (1+s)^2 - sa^2\vphat^2 +
\tfrac{a^4}{2}\sum_{k<l}\phat_k^2\phat_l^2\,,
\end{equation}
one can easily show that the eigenvalues are
bounded from below by $(1-|s|)^2$ for $|s|<1$ and $c\ge 1$. In
particular, the constraint on $c$ ensures that the combination $q\,a^2\phat_0^2$
is always positive.

Adopting the argument in ref. \cite{Hernandez:1998et}, using expansion in
Legendre polynomials,
we conclude that in the free theory the locality of the Dirac operator
\eqref{sf_neuberger} is guaranteed for this range of parameter values.

The eigenfunctions \eqref{c_one} (or \eqref{eigenfunctions}) may be
orthonormalised and
used to write down an analytical expression for the kernel $D(x,y)$ of the
Dirac operator. But the evaluation of $D(x,y)$ in this way would be
very expensive, since it involves a sum over momenta $p_0$ which in turn
are determined for each set of parameter values by the roots of
a transcendental equation.

In our computation we used a different approach. First the operator
\eqref{sf_neuberger} is half-Fourier-transformed to its time-momentum
representation $\tilde{D}(x_0,\vp)$. The square-root of the remaining
$(d+1)(T-1)\times(d+1)(T-1)$-matrices (one for each value of $\vp$) is computed
with the min-max polynomial as explained in section 4 of \cite{Giusti:2002sm}.
For the final step, to obtain the propagator, we used the built-in inversion
routine of MATLAB.

The so calculated modified Neuberger-Dirac operator has been inserted in the
Ginsparg-Wilson relation to compute the difference $\Delta_B$ and to confirm,
that it is in fact localised at the boundaries with exponentially decaying
tails.

\section{Chiral Gross-Neveu model}\label{cgn}
To test the operator \eqref{sf_neuberger} beyond free fermions we introduce
a two-dimensional theory with four-fermion interactions. The euclidean continuum
action of the chiral Gross-Neveu model can be given in the form
\begin{equation}\label{cont_action}
      S^c_{\text{CGN}} = \int \mathrm{d}x^2\,\li\{ \psibar\,\gmu\dmu\,\psi -
\tfrac{1}{2} g^2 (O_{SS} - O_{PP}) - \tfrac{1}{2} g_V^2 O_{VV} \re\}\,,
\end{equation} 
\begin{equation}
      O_{SS} =(\psibar\psi)^2\,,\quad O_{PP}
=(\psibar\gfive\psi)^2\,,\quad O_{VV} =\sum_\mu(\psibar\gmu\psi)^2\,.
\end{equation}
All Dirac and flavour indices are suppressed and contracted in a
straightforward way. For $N$ flavours of fermions this action possesses an
$U(N)$-flavour symmetry and an $U(1)$ continuous chiral symmetry. For a detailed
derivation of the possible terms (in terms of renormalisability) see chapter
5 of ref. \cite{Leder:2007aq}. The terms in \eqref{cont_action} are a full set
of allowed terms that respect the above mentioned symmetries (plus euclidean
symmetry).
For $N\ge 2$ the model shares with QCD the property of an asymptotically free
coupling (namely $g^2$).

Discretisation with Wilson fermions is straightforward. But since the Wilson
term breaks chiral symmetry a mass term and an additional coupling have to be
added
\begin{equation}\label{w_action}
   S_{\text{CGN,W}} = a^2\sum_x\,\big\{ \psibar\,(\Dw + m_0)\,\psi
                        -\tfrac{1}{2} g_w^2 (\Oss -\Opp)
                        - \tfrac{1}{2} \delta_P^2 \Opp
                         - \tfrac{1}{2} g_{V,w}^2\Ovv  \big\}\,.
\end{equation}

No additional coupling (or mass term) is necessary if one uses a Dirac
operator that solves the Ginsparg-Wilson relation. Since such an operator comes
with a lattice chiral symmetry \cite{Luscher:1998pqa}, there are no other
allowed dimension 2 operators. However, to make the action manifestly invariant
under that symmetry, one has to add
irrelevant dimension 3 operators via the substitution
$\psi\to\psihat=(1-\tfrac{a}{2}D)\psi$ in the four-fermion interaction terms.
The action then reads
\begin{equation}\label{gw_action}
   S_{\text{CGN,GW}} = a^2\sum_x\,\li\{ \psibar\,D\,\psi 
                       -\tfrac{1}{2} g_{gw}^2 (\Ohss - \Ohpp)
                       - \tfrac{1}{2} g_{V,gw}^2  \Ohvv \re\}\,.
\end{equation}

Defining now the SF of the Gross-Neveu model is straightforward. Since it is 
asymptotically free (for $N\ge 2$, which we assume from now on), the scaling
dimension of local fields is equal to their engineering dimension. Therefore
the argument about the naturalness of the SF boundary conditions in QCD
\cite{Luscher:2006df} holds also in the chiral Gross-Neveu model. And in
\cite{Leder:2007aq} it is shown that just one additional renormalisation factor
for the boundary fields has to be added.

\section{Chiral Ward identity}\label{cwi}
The continuum chiral Ward identity
\begin{equation}\label{cont_wi}
   \ev{\dmu A_\mu(x)\,\Op(y)} = 2m\ev{P(x)\,\Op(y)}\,, \quad x\neq y
\end{equation}
implies on the lattice for vanishing renormalised mass $m_R=0$
\begin{equation}\label{lat_wi}
   \<\ren{\Op}\,\md_\mu \ren{A_\mu}(x) \z = \rO(a)\,.
\end{equation}
This can be used to compute the critical value of the bare mass parameter $m_0$
and the chiral symmetry restoring value of the bare coupling $\delta_{P}$ in
\eqref{w_action} to second order in perturbation theory (PT)
\cite{Leder:2007aq}.

As indicated in \eqref{lat_wi} there are chiral symmetry breaking effects that
are lattice artifacts. In this section we compute these effects for Wilson
fermions and the proposed chiral fermions. To this end correlation functions
that match the operators in \eqref{cont_wi} and \eqref{lat_wi} have to be
defined.

Correlation functions like
\begin{equation}
   f_A(x_0) = -\frac{a^2}{2N}\,\sum_{y_1,z_1}\;
               \ev{\psibar(x)\,\go\gfive\,\psi(x)\;
                   \zetabar(y_1)\,\gfive\,\zeta(z_1)\;}
\end{equation}
correlate boundary states (here a pseudo-scalar) with current insertions (here
axial current). The boundary fields $\zeta(x_1)=P_-\psi(x)$ and
$\zetabar(x_1)=\psibar(x)P_+$ at $x_0=a$ become the non-vanishing components
of $\psi$, $\psibar$ at boundary $x_0=0$ in the continuum limit. Similarly
$f_P(x_0)$ is defined to correlate a pseudo-scalar boundary state to a
pseudo-scalar density insertion. (All correlation functions will be evaluated 
for Wilson and for chiral fermions. In the later case the fermion fields, bulk
and boundary, are substituted as indicated above eq. \eqref{gw_action}.)

With the help of these two correlation functions the bare current mass can be
computed in PT
\begin{equation}
   r(x_0) = \frac{a\tilde{\partial}_0 f_A(x_0)}{2 f_P(x_0)} = r_0(x_0) +
               r_1(x_0)g^2 + \rO(g^4)
\end{equation}
If the renormalised mass is set to zero ($m_R=0$) by demanding
$L\tilde{\partial}_0 \ren{f_A(x_0)}=\rO(a/L)$ for all $x_0$ and $\theta$, this
quantity is a direct measure of the lattice artifacts on the right hand side
of \eqref{lat_wi}.

\begin{figure}
   \centering
   \includegraphics[scale=0.6]{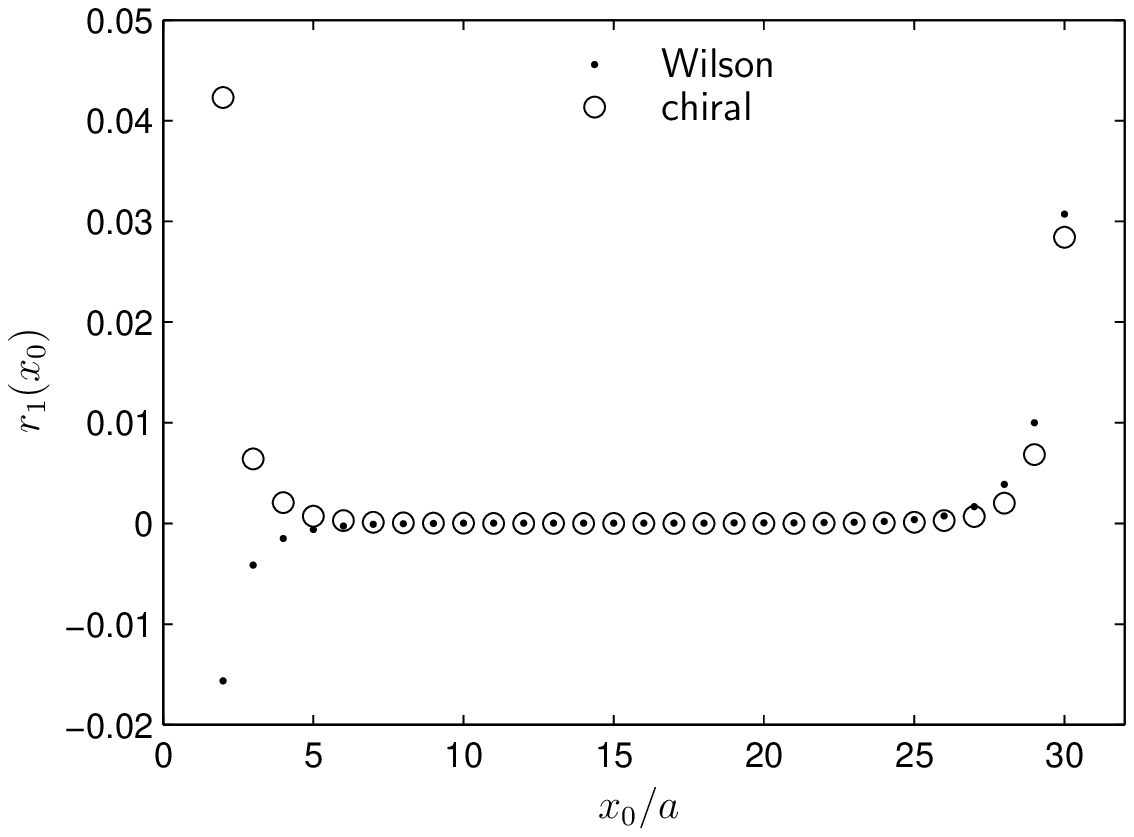}\hspace{1em}
   \includegraphics[scale=0.6]{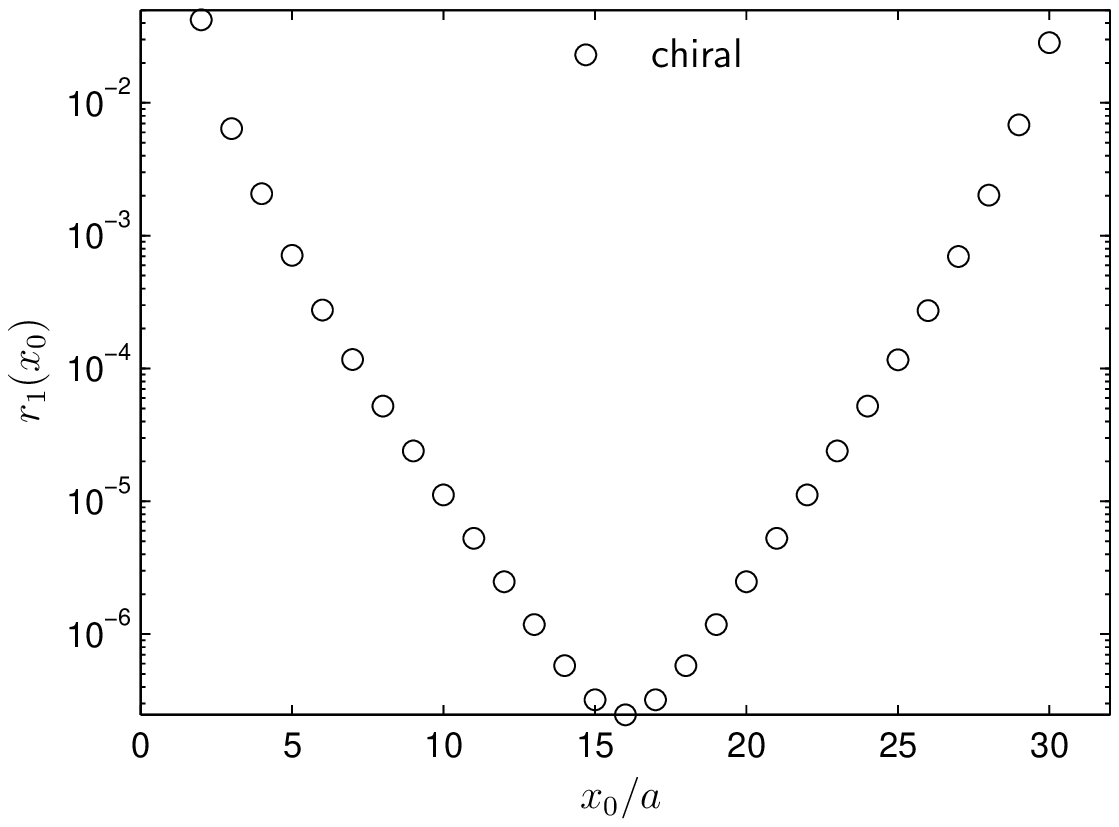}
   \caption{Bare current mass at 1-loop in PT (left). Logarithmic plot of       
            this quantity for chiral fermions (right). The plots are for a
            $16\times32$ lattice, $\theta=0$, $N=2$ and $g_V^2=0$.}
   \label{mpcac1loop}
\end{figure}

At tree-level $r_0(x_0)$ is $\rO((a/L)^2)$ for all $x_0$ for
Wilson and for chiral fermions. At 1-loop the behaviour is similar
for both discretisaions. There are chiral symmetry breaking
effects localised at the boundaries, that survive the continuum limit and decay
exponentially with the distance to the boundaries, see fig. \ref{mpcac1loop}.

\section{Renormalised coupling}\label{ren_coupling}
Having confirmed the desired chiral properties of the proposed chiral
fermions in the SF, we now renormalise the couplings at vanishing renormalised
mass and define universal quantities that have a finite and unique continuum
limit.

To this end we define boundary to boundary correlation functions
\begin{equation}
   f_4 = -\tfrac{a^4}{2(N^2-1)L^2} \sum_{u_1 v_1 y_1 z_1}\, \ev{
      \zetabar'(u_1)\, \gfive \lambda^a \zeta'(v_1)\; \zetabar(y_1)\,
      \gfive\,\lambda^a \zeta(z_1) }
\end{equation}
\begin{equation}
   f_2 = -\tfrac{a^2}{NL} \sum_{u_1 z_1}\, \ev{ \zetabar(u_1)\, \zeta'(z_1)
         }\,,\quad \ren{f_2} = Z_\zeta^2\,f_2 \quad \text{and} \quad \ren{f_4} =
         Z_\zeta^4\,f_4
\end{equation}
and the renormalisation factor free ratio
\begin{equation}
   R(\theta) = \frac{\ren{f_4}}{\ren{f_2}^2} - 1= \frac{f_4}{\li(f_2\re)^2}  - 1
            \quad \text{at}\quad m_R=0
\end{equation}
which depends on the periodicity angle $\theta$ in the spatial directions.
The $\lambda^a$s are the generators of
the algebra of $SU(N)$ and $\zeta'$, $\zetabar'$ are the equivalents of $\zeta$,
$\zetabar$ at $x_0=T-a$.

Renormalised couplings can now be defined (up to normalisation factors) as
the difference of $R(\theta)$ at two values of $\theta$ 
\begin{equation}
   \tilde{g}^2 \propto R(\theta_1)-R(\theta_2)
\end{equation}
and similar for $\tilde{g}_V^2$ \cite{Leder:2007aq}. (The normalisation factors
are functions of $\theta_1$ and $\theta_2$ only and chosen such that
 in the continuum limit $\tilde{g}^2 \sim g^2$ for small $g^2$.)
For definiteness we fix one of them to $\theta_1=0$. There are then
renormalised couplings $\tilde{g}^2_\eta$, $\tilde{g}_{V,\eta}^2$ for each
value $\theta_2=\eta$. They have an expansion in PT and the 1-loop result for
$\tilde{g}_\eta^2$ with the proper normalisation is
\begin{equation}
   \tilde{g}^2_\eta  \eqw{a\to 0} g^2_{w} +  g^4_{w}\,b_0\,\ln(a/L) +
         c_g^{w}(\eta) + \rO(g^6)
\end{equation}
for Wilson fermions and
\begin{equation}
   \tilde{g}^2_\eta  \eqw{a\to 0} g^2_{gw} +  g^4_{gw}\,b_0\,\ln(a/L) +
            c_g^{gw}(\eta) + \rO(g^6)\,,
\end{equation}
for chiral (gw) fermions. In this formulae, $b_0=-N/\pi$, is the
correct universal first coefficient of the beta-function (showing asymptotic
freedom), which was computed earlier in the continuum \cite{Mitter:1974cy}. The
finite parts found in the two computations differ as expected. The result for
the renormalised couplings $\tilde{g}_V^2$, corresponding to the vector-vector
interaction, can be found in ref. \cite{Leder:2007aq}. We here only note,
that after rearranging the terms in the action with the help of
Fierz-transformations, this coupling has (up to 1-loop) a vanishing
beta-function.

An universal renormalisation group invariant (RGI) quantity can now be
defined as the difference of the renormalised coupling for two different values
of $\eta$. This is because in the continuum a non-zero $\theta$ in \eqref{bc_3}
shifts the momenta of the external legs, but does not effect the loop integrals.
The difference
\begin{equation}
   F_{\mathrm{RGI}}(\eta,a/L) = \tilde{g}^2_\eta(a/L) -
         \tilde{g}^2_{\eta_0}(a/L)\,, \quad   \eta_0=1\,.
\end{equation}
has an expansion in the renormalised couplings (we
omit the subscript $\eta$)
\begin{equation}
   F_{\mathrm{RGI}}(\eta,a/L) = F^{(1)}_1(\eta,a/L)\, \tilde{g}^4 +
      F^{(1)}_2(\eta,a/L)\, \tilde{g}^2\,\tilde{g}^2_V + 
      F^{(1)}_3(\eta,a/L)\, \tilde{g}_V^4 +
      \rO(\tilde{g}^6)\,,
\end{equation}
where the coefficients are given by the finite parts. Since it is an universal
quantity, these coefficients have a finite and unique continuum limit
independent of the discretisation.
\begin{table}[htp]
 \centering
   \begin{tabular}{|l|l|l|l|l|l|}\hline
      $\eta$ & 0.1 & 0.2 & 0.3 & 0.4 & 0.5 \\\hline 
      $F^{(1)}_{1,\mathrm{w}}(\eta,0)$ & 5.0797(6) & 4.4704(5) & 3.6395(3) &
2.7629(1) &      1.9672(1) \\ \hline
      $F^{(1)}_{1,\mathrm{gw}}(\eta,0)$ & 5.078(2) & 4.469(2) & 3.638(2) &
2.762(2) &      1.966(2)\\\hline
   \end{tabular}
   \caption{First order coefficients in the expansion of $F_{\mathrm{RGI}}$
            with Wilson (w) and chiral            (gw) fermions.}
   \label{universal}
\end{table}
The continuum values for different choices of $\eta$ are given in tab.
\ref{universal}. The errors are estimated by fitting the first few terms of the
Symanzik expansion of lattice diagrams to the data at lattice sizes from
$4\times8$ to $64\times128$ ($128\times256$) for chiral (Wilson) fermions
(method explained in appendix D of \cite{Bode:1999sm}). The continuum values
agree within the error, thus confirming universality for the operator
\eqref{sf_neuberger}.

\section{Final remarks}\label{concl}
The formulation of the SF for fermionic models of the Gross-Neveu type in
\cite{Leder:2007aq} has enabled the first test of the recently proposed chiral
Dirac operator in the presence of the SF boundary conditions. At 1-loop we have
shown that a chiral Ward identity takes its continuum value far away from the
boundaries. The calculation of an universal quantity gives the right
discretisation independent value.

The size of the lattice artifacts has not been discussed so far. What we
observe is that the chiral fermions are tree-level $\rO(a)$-improved, if the
value of $c$ is tuned correctly. At 1-loop it is not enough to tune
$c$, because there are dimension 3 operators at the boundary, that spoil
automatic $\rO(a)$-improvement. This however is peculiar to the studied kind of
models with four-fermion interactions.

As next step, it would be desirable to perform a similar (perturbative)
computation in QCD in order to provide guidance for the final non-perturbative
application of chiral fermions in the Schrödinger functional.

\paragraph{Acknowledgement}
We would like to thank Rainer Sommer for critical reading an early version of
the text. This work was supported by Deutsche Forschungsgemeinschaft in form of
Sonderforschungsbereich SFB TR 09.

\end{document}